\documentclass[12pt]{iopart}
\usepackage{color}
\usepackage{mathbbol}              
\usepackage{graphics,graphicx,epsfig,ulem}

\begin{document}

\title{Enhanced electric field sensitivity of rf-dressed Rydberg dark states}
\author{M. G. Bason,$^1$ M. Tanasittikosol,$^1$ A. Sargsyan,$^2$ \\ A. K. Mohapatra,$^3$ D. Sarkisyan,$^2$ R. M. Potvliege,$^1$  and \\ C. S. Adams$^1$}
\address{\mbox{}$^1$ Department of Physics, Durham University,
Rochester Building, South Road, Durham DH1 3LE, UK}
\address{\mbox{}$^2$ Institute for Physical Research, Armenia National Academy of Science, 0203 Ashtarak-2, Armenia}
\address{\mbox{}$^3$ 5.\ Physikalisches Institut, Universit\"at Stuttgart, Pfaffenwaldring 57,
70569 Stuttgart, Germany}

\begin{abstract}
Optical detection of Rydberg states using electromagnetically induced transparency (EIT) enables continuous measurement of electric fields in a confined geometry. In this paper, we demonstrate the formation of rf-dressed EIT resonances in a thermal
Rb vapour and show that such states exhibit enhanced sensitivity to dc electric
fields compared to their bare counterparts. Fitting the corresponding 
EIT profile enables precise measurements of the dc field independent
of laser frequency fluctuations. Our results indicate that space
charges within the enclosed cell reduce electric field inhomogeneities within the interaction region. 
\end{abstract}

\maketitle

\section{Introduction}

Rydberg atoms are interesting in the context of precision measurement and sensing due to
their strong interatomic interactions and extreme sensitivity
to electric fields~\cite{gallagher}. The application of Rydberg states as electric field sensors has been explored using a supersonic beam of krypton atoms with principal
quantum number $n=91$ \cite{ost99}. This experiment demonstrates the potential for electrometry using highly excited Rydberg states with a measured sensitivity of
20~$\mu$V cm$^{-1}$.
However, a disadvantage of using Rydberg atoms is that the standard detection technique of pulsed field ionization means that the measurement cannot be performed continuously, and typically the detection system is relatively large.
Recently, the coherent optical detection of Rydberg states
using electromagnetically induced transparency (EIT)
has been demonstrated in a thermal vapour cell~\cite{moha07},
in an atomic beam~\cite{maug07}
and in ultra-cold atoms~\cite{weat08}. This detection technique has the advantage that it is continuous and can be performed in confined geometries down to the micron scale \cite{kueb09}. Consequently one can envisage a compact electrometry device analogous to the chip scale atomic magnetometer \cite{schw04}.

Ladder EIT \cite{eit_review} involving a Rydberg state arises due to the formation of Rydberg dark states, which are coherent superpositions of the ground state and a Rydberg state. Rydberg dark resonances result in a narrow feature in the susceptibility, and thereby an enhanced electro-optic effect compared to bare Rydberg states~\cite{moha08}. This electro-optic effect can be measured either directly,
by the displacement of the EIT dips in the absorption
spectrum~\cite{moha07}, or indirectly, by the 
phase shift of the probe field due to the electric-field dependence
of the refractive index ~\cite{moha08}. By reducing the probe laser intensity the Rydberg population in the dark state can be made vanishingly small while the change in transmission remains unchanged. Consequently, the effect of collisional ionization,
which is often observed in high density vapours \cite{keel08},
can be eliminated.

Possible applications of Rydberg dark states include single-photon entanglement~\cite{fried05}, the generation of exotic entangled states \cite{moll08} and mesoscopic quantum gates~\cite{mull09}. In addition they are of interest to applications in electrometry, owing to their giant dc Kerr coefficient~\cite{moha08}.
However, the sensitivity of a Rydberg dark state electrometer is limited by laser frequency fluctuations. Since reducing those to a suitable level entails a considerable experimental overhead,
a technique to measure electric fields that is insensitive to the absolute laser frequency is desirable.
Measurements of the splitting between D Stark sublevels \cite{moha07}
or between Stark states of higher angular momentum may
be used to this end.

In this paper, we demonstrate the formation of
Rydberg dark states dressed by a radio frequency (rf) field.
Microwave or rf dressing of Rydberg states has previously been observed
using laser excitation and field ionization of an
atomic beam \cite{bayf81,zhang94} or of cold atoms \cite{ditz09}. Here, however,
the resulting Floquet states are observed as EIT resonances
in the absorption spectrum of the probe laser beam in a vapour cell. 
We show that these rf-dressed dark states have an enhanced 
sensitivity to dc electric fields. We also show that the strength
of the dc electric field 
is encoded not only in the overall shift of the corresponding
EIT feature but also in the 
shape of the transparency window.
As a consequence, and as we illustrate by an actual measurement, rf dressing may
help determine
the dc field without absolute knowledge of the laser 
frequency.
 In addition, we show how the line shape of the dark state resonance provides information about the electric field inhomogeneity within the interaction region.

\section{Theory}

We focus on EIT in the ladder system in which a weak probe laser (wavelength 780~nm) is scanned through resonance with the transitions from
the ground 5S$_{1/2},F=2$ state to the 5P$_{3/2},F'=(2,3)$
states of $^{87}$Rb, in the presence of both dc and ac electric fields and a coupling laser (wavelength 480~nm) resonant with
the 5P$_{3/2},F'=3$ to the 32S$_{1/2},F''=2$ transition.
Such a ladder system is ideal for rf dressing as the two components of the dark state have a large
differential shift in an external field and there is no additional splitting of the $J = 1/2$ Rydberg state.

At any point in the cell, the local applied electric field can be written as
${\cal E}(t)={\cal E}_{\rm dc} + {\cal E}_{\rm ac} \sin\omega_{\rm m}t$.
In the experiment,
$\omega_{\rm m}/2\pi$ ranges from 10 to 30 MHz, which is much less than
the relevant optical transition frequencies.
Ignoring the laser fields and decoherence for the time being,
we can work within the adiabatic approximation and describe each of the
hyperfine components of the
32S$_{1/2}$ state
by a time-dependent state vector of the form \cite{Pont}
\begin{equation}
|\Psi(t)\rangle
=\exp\left(-{{\rm i}\over \hbar}\int^t E[{\cal E}(t')] \, {\rm d}t'\right)
|\psi[{\cal E}(t)]\rangle~.
\label{eq:Psi}
\end{equation}
In this expression,
$|\psi({\cal E})\rangle$ is the Stark state that develops from the
field-free state when
a static field is adiabatically turned on from 0 to ${\cal E}$,
and $E({\cal E})$ is the corresponding eigenenergy of the Stark Hamiltonian.
The dc and ac fields are sufficiently weak that
one can take
$E({\cal E})= E^{(0)} - (\alpha/2) {\cal E}^2$ and
$|\psi({\cal E})\rangle = |\psi^{(0)}\rangle+ {\cal E}|\psi^{(1)}\rangle$,
where $\alpha$ denotes the static dipole polarizability of the Rydberg state
\cite{Stoicheff},
$E^{(0)}$ and $|\psi^{(0)}\rangle$ are the energy and state vector in the absence of the
field, and $|\psi^{(1)}\rangle$ is
the first-order coefficient of the perturbative expansion of the Stark state
$|\psi({\cal E})\rangle$ in powers of ${\cal E}$.
Since the field
${\cal E}(t)$ is monochromatic, the state vector
of the rf-dressed Rydberg state can also be written in the Floquet-Fourier
form
\begin{equation}
|\Psi(t)\rangle = \sum_{N=-\infty}^{\infty}
\exp(-{\rm i}\epsilon_N t/\hbar)
|\psi_N\rangle,
\label{eq:Floquet}
\end{equation}
with $\epsilon_N = \epsilon_0 + N\hbar\omega_{\rm m}$. 
A short calculation \cite{zhang94,ditz09} shows that
within the approximations made above,
\begin{equation}
\epsilon_N    = 
E^{(0)} - {\alpha^2 \over 2}
{\cal E}^2_{\rm dc} - {\alpha^2 \over 4}{\cal E}^2_{\rm ac} + N\hbar\omega_{\rm m}
\label{eq:epsilon} 
\end{equation}
and
$|\psi_N\rangle=B_N|\psi^{(0)}\rangle+
C_N|\psi^{(1)}\rangle$, with
\begin{equation}
B_N=\sum_{M=-\infty}^\infty
{\rm i}^{N-2M}J_{2M-N}\left({\alpha {\cal E}_{\rm dc}{\cal E}_{\rm ac}\over \hbar\omega_{\rm m}}\right)
J_M\left({\alpha {\cal E}^2_{\rm ac}\over 8\hbar\omega_{\rm m}}\right)
\label{eq:Bexp}
\end{equation}
and $C_N = B_N{\cal E}_{\rm dc}+
(B_{N+1} - B_{N-1}){\cal E}_{\rm ac}/(2 {\rm i})$.
Hence,
when decoherence is ignored, each of the two hyperfine components of the
32S$_{1/2}$ state effectively turns
into a manifold of equally spaced states
under the action of the ac field \cite{bayf81,zhang94,ditz09}. 
Because the vectors $|\psi^{(0)}\rangle$ and
$|\psi^{(1)}\rangle$ have opposite parity, $|\psi^{(1)}\rangle$ does not
couple to the 5P$_{3/2}$ state. Therefore,
the Rabi frequency for the
transition from a particular hyperfine
component of this state to a particular
component of the dressed Rydberg state
differs from the corresponding zero-field Rabi frequency only
by a factor $|B_N|$ (for the transition to the component with energy
$\epsilon_N$). The 5S$_{1/2}$ and 5P$_{3/2}$ states are also 
dressed by the applied field, but their 
polarizability is too small for any of
their Floquet sideband states to be significantly
populated at the ac field amplitudes considered here.

It follows that under the action
of the ac field, and provided the relevant relaxation times are much longer
than $2\pi/\omega_{\rm m}$,
one should expect that the Rydberg
dark states turn into Floquet manifolds of dark states
and that each EIT dip in the absorption spectrum
acquires multiple side bands.
That such Floquet dark states can be obtained
in conditions easily accessible to experiment
is one of the results of this paper.
(In the opposite limit where the decoherence time is much shorter than
the period of modulation,
EIT happens as if the applied electric field is
static and the experimental signal is the time-average of the
instantaneous absorption
spectrum over the distribution of values of ${\cal E}(t)$.)

\section{Comparison between theory and experiment} 

\begin{figure}[t!]
\begin{center}
\includegraphics[width=8.0cm]{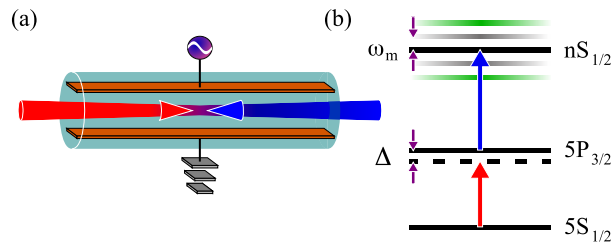}
\caption{
(a) Schematic of the experimental set up. The 780~nm probe beam (red) and 480~nm coupling beam (blue) counterpropagate through a Rb cell containing parallel plane electrodes. The dc and the ac fields are applied by imposing a potential difference across the electrodes. (b) Schematic of the energy levels scheme. The detuning $\Delta$ of the probe beam is varied. The coupling beam is resonant with the transition between the intermediate 5P$_{3/2},F'=3$ state and the Rydberg nS$_{1/2},F''=2$ state (n $=32$). An applied electric field with angular frequency $\omega_{\rm m}$ generates a ladder Floquet state separated by integer multiples of $\omega_{\rm m}$. The first order Floquet dark states (grey) are particularly sensitive to any dc field.
}
\end{center}
\label{fig1}
\end{figure}

Schematics of the experimental set up and of the energy levels scheme are
shown in figure 1. We employ a specially fabricated 11~mm-long Rb vapour cell containing two
parallel plane electrodes running along its whole length and separated by a 5 mm gap. The probe beam and the co-axial, counter-propagating coupling beam are directed along the electrode cell axis and are both polarized parallel to the electrodes. Each beam is focussed using 10~cm lenses. The probe beam has an input power of 300~nW and a $1/\textrm{e}^2$ radius of 1.7~mm. The corresponding values for the coupling beam are 40~mW and 1.0~mm. The latter is stabilized against slow drift using EIT in a reference cell~\cite{abel09}. 
The transmission through the electrode cell is monitored as a function of the probe detuning. To increase the number density of Rb atoms the electrode cell is heated to c.\ 40 $^\circ$C.  The probe beam is split into two with one component propagating through the electrode cell, and the other passing through a longer room temperature cell. By subtracting these two signals, the Doppler background is removed. Measuring the off-resonant probe beam power after the electrode cell allows the change in transmission, $\Delta \mathrm{T}$, to be calibrated. The detuning axis is calibrated using saturation/hyperfine pumping spectroscopy.
    
A typical EIT spectrum with an applied ac field
is shown in figure 2 together with theoretical fits for different spatial
profiles of the electric field. 
The theoretical model of EIT is based on the steady-state solution
of the optical Bloch equations in the weak probe limit~\cite{gea95},
assuming that the different magnetic sub-levels of the ground state
are equally populated. In the experiment, the intensity of the probe beam, $I \approx 3 I_{\rm sat}$, is considerably higher
than the ideal weak probe limit~\cite{sidd08}, however, good agreement can still be obtained.
The absorption spectrum is obtained by integrating the
absorption coefficient 
over the velocity distribution of the atoms,
the spatial profiles of the two laser beams and the length of
the cell.
The absorption coefficient is a function of four parameters
that must be derived from the data,
namely the Rabi frequency for the
5P$_{3/2},F'=3$ to 32S$_{1/2},F''=2$ transition,
the dephasing rates of the 5P$_{3/2}$--5S$_{1/2}$ and
32S$_{1/2}$--5P$_{3/2}$ coherences, and the temperature of the vapour. The
values of these experimental
parameters are obtained by a least-square fit of the
theoretical change in transmission, $\Delta \mathrm{T}$, to the data. The fit covers a range of probe detunings encompassing both 
the 5P$_{3/2},F'=2$ and $F'=3$ states whereas only latter is shown in figure 2.
The mean separation between the plates is not known with sufficient 
precision, and therefore is derived from the data
by fitting the Stark shift
of the EIT resonances to equation (\ref{eq:epsilon}) for a number of different
values of the applied voltage.

\begin{figure}[t!]
\begin{center}
\includegraphics[width=8.0cm]{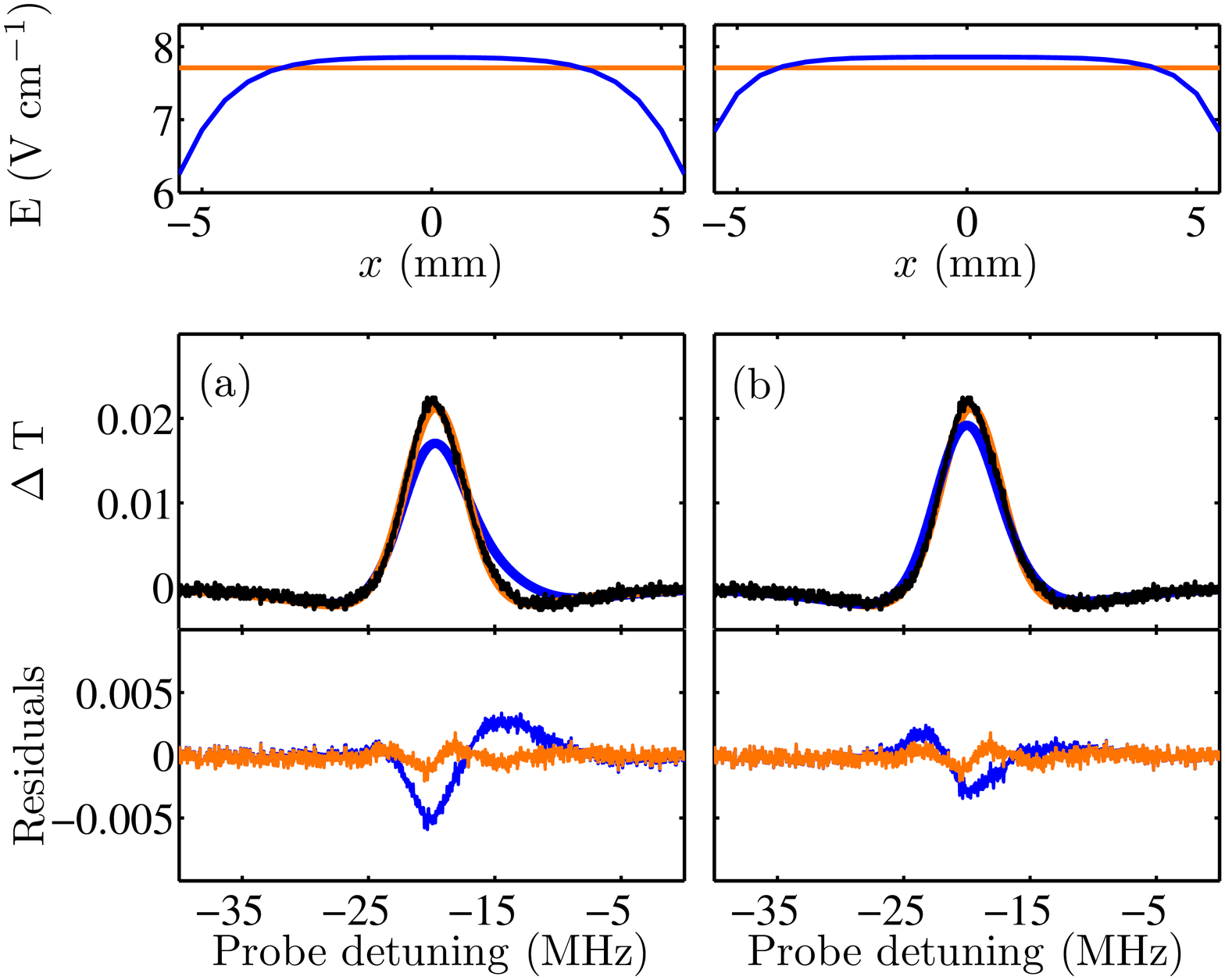}
\caption[]{
Comparison between the observed EIT spectra (black)
for an applied ac electric field, ${\cal E}_{\rm ac}=7.7$ V cm$^{-1}$,
and the theoretical model for the field calculated for the actual electrode geometry (blue) and for a uniform field (red).
In the theoretical model, it is assumed either that the laser beams are
aligned exactly on the longitudinal axis of the two
electrodes (a)
or that they are parallel to this axis but are offset by 1 mm towards
one of the plates (b).
The corresponding electric fields distributions are shown at the top.
In this example, the electric field frequency is 26~MHz.}
\end{center}
\label{fig2}
\end{figure}

The spatial profile of the applied electric field along the laser beam axis is unknown due to the possibility of free charges inside the cell \cite{moha07}. 
The field produced by the Stark plates drops significantly at their
edges (see the top panels of figure 2). If we assume that the field
experienced by the atoms exposed to the laser beams varies
accordingly, the resulting theoretical EIT profile is asymmetric and
inconsistent with the data (see the middle and bottom panels of figure 2). The theoretical results are found to be in good agreement with the data if the
field is assumed to be uniform. This would occur if the free charges inside the
cell equalize the electric field in the interaction region.
We note in this respect that charges can be created by
the photoelectric effect where the coupling laser intersects
the Rb vapour on the inner surface of the cell, the photoelectrons being
ejected from the surface. A net positive charge
of $10^6 e$ on each window distributed over the waist of the coupling laser
is sufficient to produce
a total field with
a spatial profile consistent with the experimentally observed lineshape. We cannot exclude a misalignment of
the laser beams with respect to the centre of the plates, but the
resulting offset would be less than 1 mm and even at 1 mm 
from the axis the field produced by the plates is more inhomogeneous than is compatible with the data (see the right-hand column of figure 2).

\section{Observation of Floquet dark states}

Following the analysis in the previous section we assume that the electric field experienced by the atoms is homogeneous, 
and measure spectra for different combinations of ac and dc applied fields (${\cal E}_{\rm dc}$ and ${\cal E}_{\rm ac}$). The results are presented in figure 3. The theoretical spectra
are calculated assuming that the Rydberg state is described by the Floquet state vector (\ref{eq:Floquet})
and that the four experimental parameters mentioned in Section 3
have the same values as in the zero-field case.

In figure 3(b), we consider a case where only a pure ac field is applied.
As compared to figure 3(a), the main EIT peak
is shifted by the ac Stark effect
and the EIT profile acquires sidebands
at the second harmonics of the modulation frequency. The spacing between the sidebands and the carrier in the transmission spectrum is smaller than $2 \omega_{\rm m}/(2\pi)$ by a factor 480/780 due to the Doppler mismatch. In view of the good agreement between theory and the measured data, we attribute the observed sidebands 
to the formation of Floquet dark states.

There are no odd-order sidebands in the absence of a dc field since
$B_N = 0$ for odd values of $N$ when ${\cal E}_{\rm dc} = 0$.
Their absence is consistent with the dipole
selection rules, which forbid transitions from a P state to an S state
by exchange of one laser photon and an odd number of rf photons.
The effect of adding a weak dc offset is shown in figure 3(c).
The spectrum acquires
first order sidebands, since
$|\psi^{(0)}\rangle$ contributes to every $|\psi_n\rangle$
 when ${\cal E}_{\rm dc}\not=0$.
As can be seen by comparing figures 3(c) and 3(d), increasing the dc field
modifies the EIT spectrum in several ways.
The largest change 
occurs on the $+1$ sideband.
In contrast, the additional Stark shift in the position of the central peak,
which would be the only effect of the increase in the dc field
in the absence of the modulation,
is almost invisible on the scale of the figure.
The comparison demonstrates the enhanced dc field sensitivity of
rf-dressed Rydberg dark states.

\begin{figure}[t!]
\begin{center}
\includegraphics[width=8.0cm]{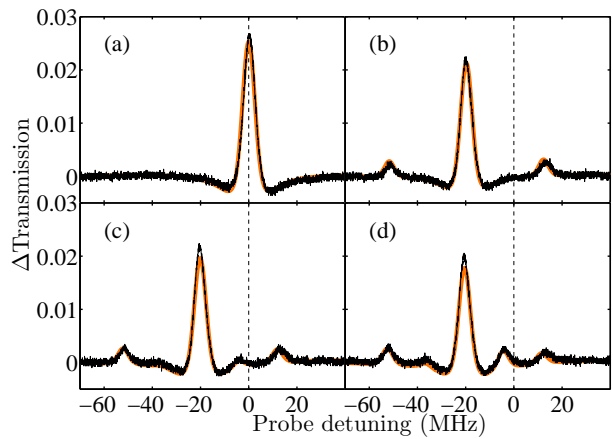}
\caption[]{
The effect of ac and dc electric fields on the Rydberg EIT spectrum.  The thick curves are theoretical predictions based on the solution of the steady state optical Bloch equations. (a) Spectrum with no applied field. (b) Spectrum with an rf field with a frequency of 26~MHz (${\cal E}_{\rm ac} = 7.7$  V cm$^{-1}$). (c) and (d) Spectrum with the rf field plus a dc field; ${\cal E}_{\rm dc} = 0.4$ V cm$^{-1}$ in (c) and 0.8 V cm$^{-1}$ in (d).
}
\end{center}
\label{fig3}
\end{figure}

\begin{figure}[t]
\begin{center}
\includegraphics[width=8cm]{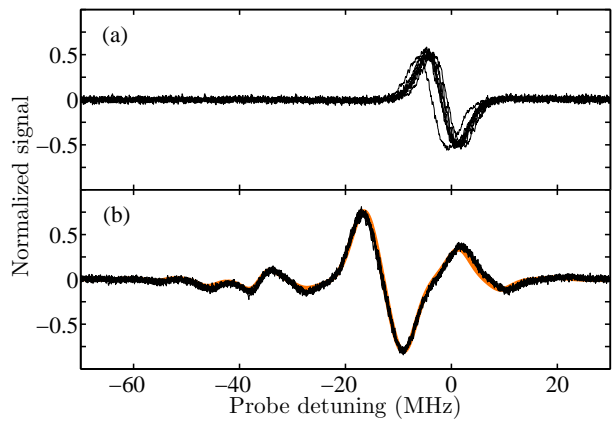}
\caption[]{
The dc field signal obtained by turning the dc field on and off at 50~kHz and using lock-in detection. The same vertical scale is used in (a) and in (b). (a) The difference signal measured for ${\cal E}_{\rm ac}=0$ and ${\cal E}_{\rm dc}= 1.37$ V cm$^{-1}$. Ten data sets are presented; the results shown are not corrected for the frequency fluctuations of the probe laser. (b) The difference signal measured for ${\cal E}_{\rm ac}= 7.14$ V cm$^{-1}$ and ${\cal E}_{\rm dc}= 1.37$ V cm$^{-1}$. The modulation frequency is 15 MHz. As in (a), ten data sets are presented, but here they are corrected for the frequency fluctuations of the probe laser using the Floquet model. The thick curve shows the Floquet EIT result calculated for a value of ${\cal E}_{\rm dc}$ ensuring an optimal fit between the model and one of the data sets.
}
\end{center}
\label{fig4}
\end{figure}

\section{Enhanced electric field sensitivity of Floquet dark states}

To further illustrate this enhancement effect,
we now show that adding an ac modulation helps deduce the dc component
of the electric field from the EIT resonance.
The difference between transmission spectra measured with
and without applying a dc field are shown in figure 4. Here we
employ a double modulation technique where the dc field is switched
on and off at a frequency of 50 kHz, i.e., much less than the modulation frequency. 
The effect of the dc field is then extracted using lock-in detection
resulting in a derivative lineshape.
Each thin black line represented
in figure 4 gives the difference signal averaged over four consecutive
scans. Figure 4(a) shows the difference spectrum in the absence of ac field
for 10 individual data sets.
The spread in the data reflects the instability of the probe laser.
Deriving ${\cal E}_{\rm dc}$ from these results is hindered by the
fact that only the position of the EIT feature on the frequency axis, and
not its shape, varies significantly with
the strength of the dc field.
Due to the experimental uncertainty in the probe frequency, a value of
${\cal E}_{\rm dc}$ cannot be obtained by fitting the model to the data from
figure 4(a). However, the fit is possible
when these data are augmented by the difference
EIT signal arising from the
5P$_{3/2},F'=2$ state and by frequency calibration data obtained by
saturation spectroscopy.
The theoretical difference signal is calculated using the same values of the
Rabi frequency, dephasing rates and temperature as in figure 3(a). We
treat ${\cal E}_{\rm dc}$ as an unknown parameter.
For comparison with the theory, we correct the experimental results for
random variations in the calibration of the probe frequency and of the signal
by rescaling and shifting the origins of
the respective axes. The corresponding offsets
and scaling factors are found for each individual data set, together with 
${\cal E}_{\rm dc}$, by fitting the rescaled experimental difference signal 
to the model.
From the 10 values of ${\cal E}_{\rm dc}$ obtained in this way, we find that
${\cal E}_{\rm dc}=1.6\pm 0.4$ V cm$^{-1}$. This value
is in agreement with 
that derived from the 
dc voltage applied to
the electrodes, $1.37\pm 0.02$ V cm$^{-1}$, but it has a larger uncertainty.

A constant ac field is then added and the lineshape extracted once more, figure 4(b). The lock-in detection still only detects changes in signal due to the dc field. In this case, for the same
dc field, the difference signal is larger and contains more features. As the details of these features depend on ${\cal E}_{\rm dc}$,
the change in the spectrum due to the dc field is readily separated from
the frequency fluctuations of the probe laser.
Measuring ${\cal E}_{\rm dc}$ is thus easier.
The theoretical difference signal is calculated in the same way as in the
absence of the rf field. We
assume for ${\cal E}_{\rm ac}$
the value derived from the voltage applied to the electrodes.
From the 10 values of ${\cal E}_{\rm dc}$ obtained by fitting the data from 
figure 4(b) to the model, we find that
${\cal E}_{\rm dc}=1.36\pm 0.04$ V cm$^{-1}$
(or $1.40\pm 0.03$ V cm$^{-1}$
when the saturation spectroscopy data and the signal from
5P$_{3/2},F'=2$ state are also taken into account).
For these parameters, introducing an
ac modulation thus reduces the uncertainty in the dc field measurement by one order of magnitude.

In the case considered in figure 4, the application of
the ac field also increases the amplitude of the difference signal
by about 50\%. As shown in  figure 5, larger enhancements (of up to 3) can be obtained for
other combinations of dc fields and modulation frequencies.


\begin{figure}[t!]
\begin{center}
\includegraphics[width=8cm]{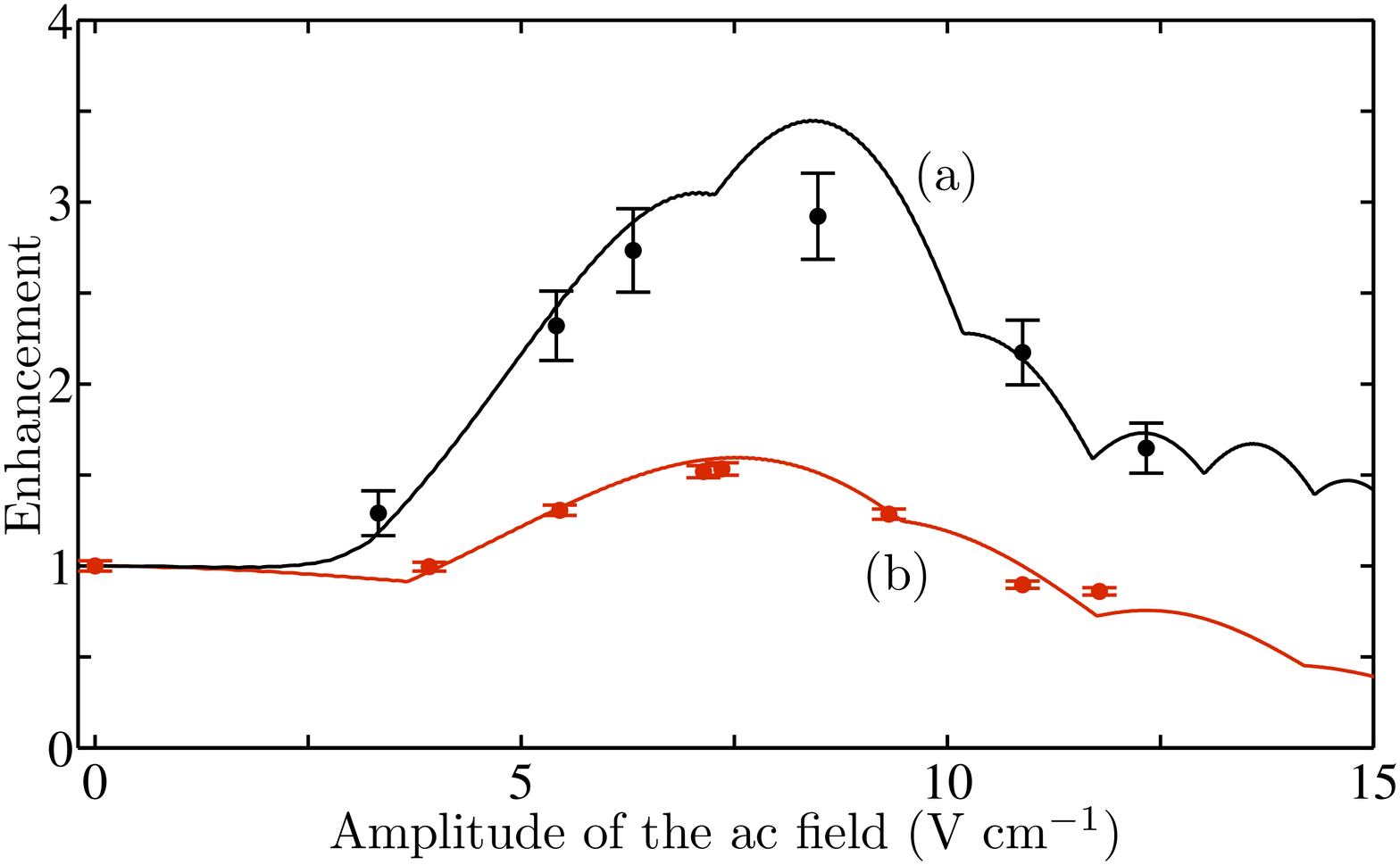}
\caption[]{
The increase in the amplitude of the dc electrometry signal as a function of ${\cal E}_{\rm ac}$, (a) for a modulation frequency of 10~MHz and ${\cal E}_{\rm dc}=  0.78$ V cm$^{-1}$, (b) for a modulation frequency of 15~MHz and ${\cal E}_{\rm dc}= 1.37$ V cm$^{-1}$. The points are experimental data and the lines are the theoretical predictions of the model presented in the text. The sudden changes in the theoretical curves occur at zeroes of the Bessel functions appearing in equation (\ref{eq:Bexp}).
}
\end{center}
\label{fig5}
\end{figure}


\section{Summary and outlook}

In summary,
we have demonstrated the formation of
Floquet dark states induced by the application of an ac field to a ladder system involving a highly polarizable Rydberg state. We have shown that these states display enhanced sensitivity to dc electric fields and provide information on the strength of the dc field independent of the laser frequency.
Potentially,
an ac modulation may thus facilitate the measurement of the local electric
field inside a vapour cell,
which is a relevant issue in the control of 
Rydberg-Rydberg interactions.
The simple theory outlined above can be generalized to the case where
the field splits the Rydberg state into several Stark components, thereby
opening the possibility of using ac modulation to
enhance the sensitivity of
measurements based on the D states or on states of
higher angular momentum.
We
have shown that charge imbalances in an enclosed vapour cell
can cancel the spatial inhomogeneities of the field, therefore for local field measurements
the interaction region may need to be limited to a small volume.
This would be the case, for instance, in a 3-photon Doppler-free excitation scheme in which the two
pump and probe laser beams intersect at appropriate angles within a restricted volume.

%
%

\section*{Acknowledgments}
We thank the EPSRC and the DPST Programme of the
Thai Government for financial support.

\end{document}